\documentclass[conference]{IEEEtran}
\IEEEoverridecommandlockouts

\usepackage{cite}
\usepackage{amsmath,amssymb,amsfonts}
\usepackage{algorithmic}
\usepackage{graphicx}
\usepackage{textcomp}
\usepackage{xcolor}
\usepackage{subfloat}
\usepackage{subfig}
\usepackage{url}

\usepackage{wrapfig}

\usepackage{lipsum}

\usepackage[utf8]{inputenc}

\usepackage{booktabs}

\usepackage[export]{adjustbox}

\usepackage{multirow}
\usepackage{multicol}

\bgroup

\def\BibTeX{{\rm B\kern-.05em{\sc i\kern-.025em b}\kern-.08em
		T\kern-.1667em\lower.7ex\hbox{E}\kern-.125emX}}

\title{Sleep Stage Classification Using Bidirectional LSTM in Wearable Multi-sensor Systems}

\DeclareRobustCommand*{\IEEEauthorrefmark}[1]{%
	\raisebox{0pt}[0pt][0pt]{\textsuperscript{\footnotesize #1}}%
}

\author{
	\IEEEauthorblockN{
		Yuezhou Zhang\IEEEauthorrefmark{1}, Zhicheng Yang\IEEEauthorrefmark{2}, Ke Lan\IEEEauthorrefmark{1}, Xiaoli Liu\IEEEauthorrefmark{3}, Zhengbo Zhang\IEEEauthorrefmark{4}$^*$\thanks{$^*$Zhengbo Zhang, Peiyao Li, and Desen Cao are with Dept. of Biomedical Engineering, and Medical Device R\&D and Evaluation Center. Zhengbo Zhang is the corresponding author.}, Peiyao Li\IEEEauthorrefmark{4}, \\ Desen Cao\IEEEauthorrefmark{4}, Jiewen Zheng\IEEEauthorrefmark{1}, Jianli Pan\IEEEauthorrefmark{5}
	}
	\IEEEauthorblockA{
		\IEEEauthorrefmark{1}Beijing SensEcho Science \& Technology Co., Ltd., Beijing, China\\
		\IEEEauthorrefmark{2}University of California, Davis, CA, USA\\ 		
		\IEEEauthorrefmark{3}Beihang University, Beijing, China\\
		\IEEEauthorrefmark{4}Chinese PLA General Hospital, Beijing, China\\
		\IEEEauthorrefmark{5}University of Missouri - St. Louis, MO, USA\\
		vincent.cheng@wearable-health.com, zhengbozhang@126.com
	}
}

\begin{document}
	
\maketitle

\begin{abstract}
Understanding the sleep quality and architecture is essential to human being's health, which is usually represented using multiple sleep stages. A standard sleep stage determination requires Electroencephalography (EEG) signals during the expensive and labor-intensive Polysomnography (PSG) test. To overcome this inconvenience, cardiorespiratory signals are proposed for the same purpose because of the easy and comfortable acquisition by simplified devices. In this paper, we leverage our low-cost wearable multi-sensor system to acquire the cardiorespiratory signals from subjects. Three novel features are designed during the feature extraction. We then apply a Bi-directional Recurrent Neural Network architecture with Long Short-term Memory (BLSTM) to predict the four-class sleep stages. Our prediction accuracy is 80.25\% on a large public dataset (417 subjects), and 80.75\% on our 32 enrolled subjects, respectively. Our results outperform the previous works which either used small data sets and had the potential over-fitting issues, or used the conventional machine learning methods on large data sets.
\end{abstract}


\section{Introduction}
\label{sec:intro}

Sleep quality is essential to human being's health. To analyze it, sleep is usually classified into multiple stages in which physiological signals have different patterns, related to various physiological functions. 
Although there are many criteria to classify the sleep stages, the most common four-class sleep stages criterion is our focus, which includes \emph{Wake}, \emph{Light sleep}, \emph{Deep sleep}, and \emph{Rapid Eye Movement (REM)} \cite{berry2012aasm}. For example, deep sleep helps body metabolism and wound healing, while dreaming occurs during the REM period most. The accurate classification of sleep stages is necessary to properly analyze one's sleep architecture, further for the sleep disorder diagnose and recovery.

Polysomnography (PSG) is a standard method for sleep quality analysis, which has been used for sleep stage classification in past decades, as it provides the electrical activities of various body parts. Electroencephalography (EEG) with the assistance of Electrooculography (EOG) and Electromyography (EMG) during the PSG test is used to determine the different sleep stages.
However, the entire procedure of sleep stage classification using PSG is expensive and labor-intensive, as it requires the dedicated equipment and laboratory as well as the clinical specialists' expertise. Rather than relying on EEG signals from PSG, recent researchers focus on using cardiorespiratory signals because they can also indicate the different sleep stages and can be captured in plenty of the low-cost simplified devices. This approach is easily adapted into the various scenarios of mobile and Internet of Things (IoT) healthcare, sleep health monitoring in smart home \cite{yang2017vital} and other accommodations \cite{dor2012experiences}.
As the deep learning methods are rapidly developing in recent years, some latest researches have demonstrated the effectiveness that applying Recurrent Neural Network (RNN) to capture the implicit patterns of the time-series cardiorespiratory signals  \cite{zhang2017sleep,zhao2017learning}.

In this paper, we present our ongoing work that uses our designed wearable multi-sensor system to capture subjects' ECG and breathing signals, and that applies a Bi-directional RNN architecture with Long Short-term Memory (BLSTM) \cite{graves2005framewise} to classify the four sleep stages. An overview of our contributions is provided below:

\begin{enumerate}
	\item We propose and demonstrate three novel features which are effective to detect the sudden variation of RR intervals\footnote{The term of \textit{RR interval} will be explained in Sec~\ref{sec:rr-intervals}.}. A total of 152 features are extracted for the learning procedure.
	\item Compared to the recent works, we leverage a large public dataset and our own dataset to achieve the outstanding sleep stage prediction in terms of both accuracy (80.25\% on the public dataset, and 80.75\% on our own dataset) and sample size (total 449 subjects). These results outperform the previous works on the four-class sleep stage classification, which either used small data sets having the potential over-fitting issues \cite{fonseca2015sleep,muzet2016assessing,zhao2017learning,hong2018noncontact}, or used the conventional machine learning methods on large data sets \cite{tataraidze2016sleep,aggarwal2018sleep}.
	\item We design our own low-power and low-cost wearable multi-sensor vest, \emph{SensEcho}, with the support of the secure cloud system. We validate that it is robust to acquire the cardiorespiratory signals compared to the standard PSG test.
\end{enumerate}

This paper is organized as follows. Related works are discussed in Section~\ref{sec:related-works}. We describe our hardware and methods in Section~\ref{sec:methods}. The evaluations are presented in Section~\ref{sec:eval}. Section~\ref{sec:discussion} proposes further potentials and future works. We finally conclude our work in Section~\ref{sec:conclusion}.

\begin{figure}[t]
	\centering
	\includegraphics[width=0.6\linewidth]{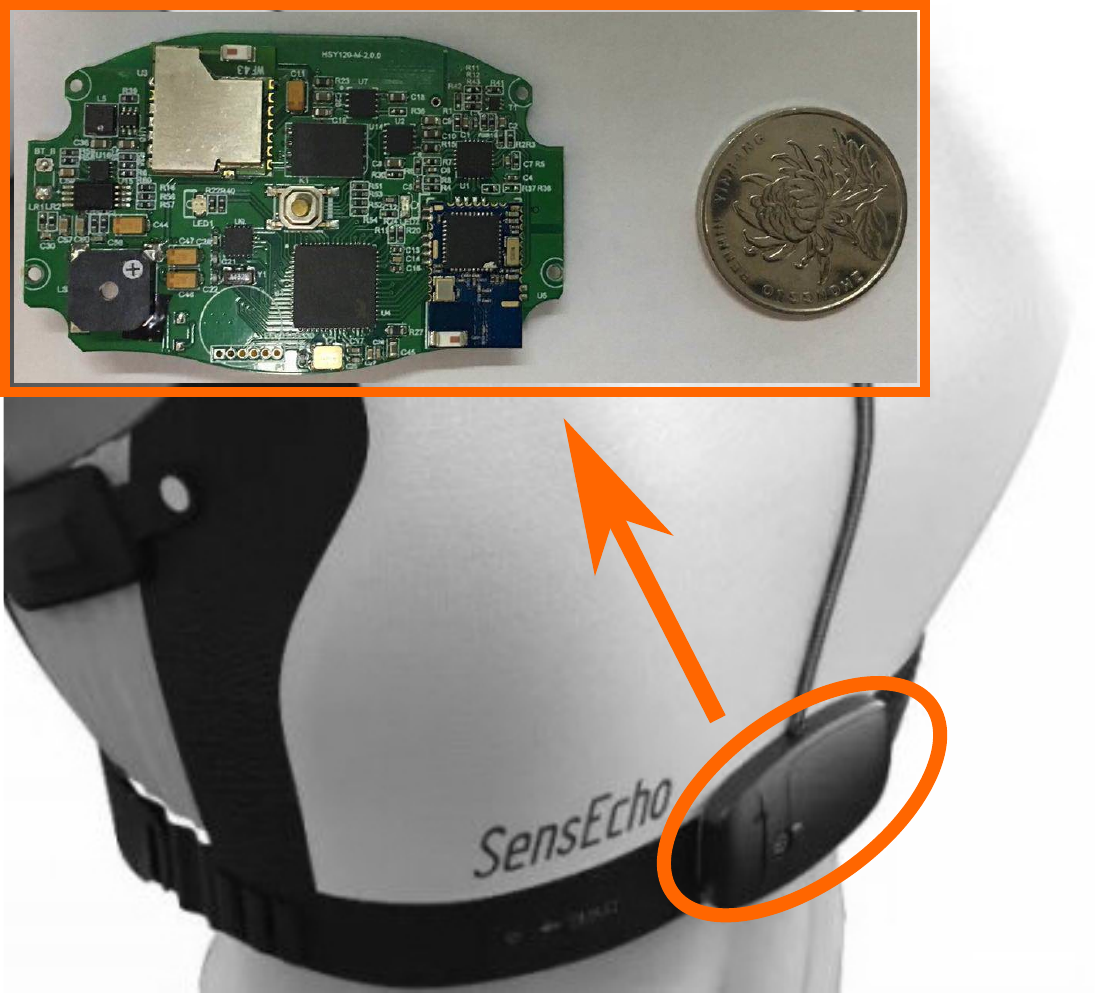}	
	\caption{\small{Hardware of SensEcho}}
	\label{fig:main-board-smaller}
	\vspace{-15pt}
\end{figure}
\section{Related Works}
\label{sec:related-works}

\subsection{EEG for Sleep Stage Classification}
Since specialists rely on EEG, EOG, EMG signals to determine sleep stages, automated classifying sleep stages based on those signals are well studied to reduce the input of labor and cost. 
Authors in \cite{pan2012transition} presented a transition-constrained discrete hidden Markov models for automatic sleep stage classification using the EEG, EOG and EMG signals of 20 subjects. 
A deep learning model, DeepSleepNet, was proposed in \cite{supratak2017deepsleepnet} for automated grading sleep stages using the single-channel EEG. DeepSleepNet consists of Convolutional Neural Networks (CNN) to extract time-invariant features and BLSTM to learn stage-to-stage transition rules from EEG signals. Again, automated classification relying on EEG signals still requires PSG tests.

\subsection{Cardiorespiratory Signals for Sleep Stage Classification}
\begin{figure}[t]
	\includegraphics[width=\linewidth]{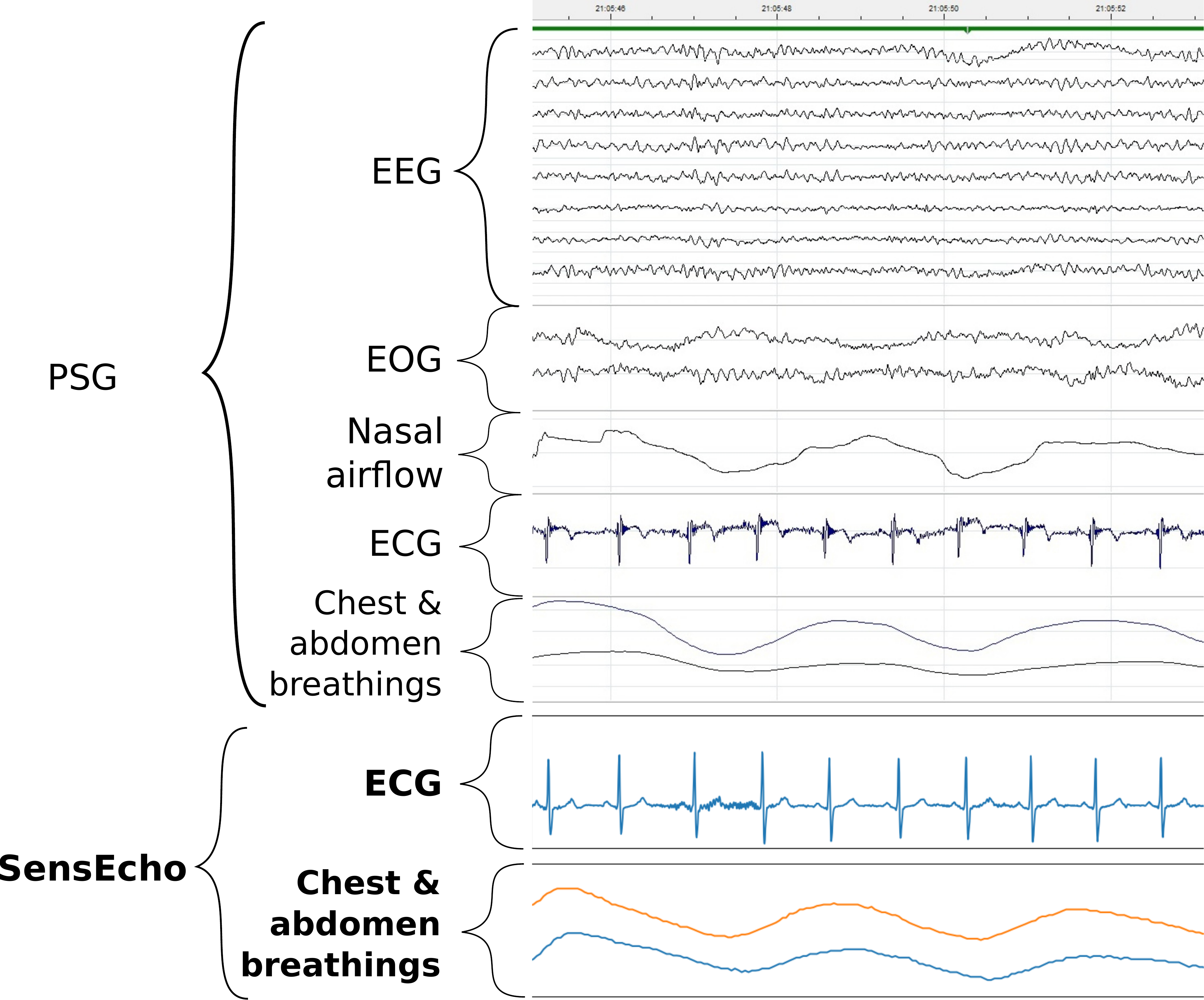}
	\caption{\small{Overview of signals from PSG and SensEcho}}
	\vspace{-5pt}
	\label{fig:signal-overview}
\end{figure}

Since it is difficult to compare the works when the number of sleep stages is not consistent, we mainly review the previous works only on the four sleep stage classification as our focus.

Large datasets \cite{tataraidze2016sleep,aggarwal2018sleep} containing the cardiorespiratory recordings were leveraged to predict the sleep stages. Due to the limitation of conventional machine learning algorithms, the prediction accuracies were moderate. Authors in \cite{fonseca2015sleep} extracted a total of 142 features from electrocardiogram and thoracic respiratory effort of 25 subjects, and applied Bayesian linear discriminant for the four-class sleep stage classification on a 30-second epoch basis. Similarly, the traditional Support Vector Machine algorithm was applied to the cardiorespiratory features of 48 subjects \cite{muzet2016assessing}. Recently, the contactless sensing methods such as using Radio Frequency (RF) signals to acquire the heart beats and breathing efforts have gained much attention. Authors in \cite{zhao2017learning} proposed a new learning framework to involve various deep learning methods on the RF signals of 25 subjects to classify sleep stages into four classes with the high prediction accuracy. The BLSTM-based five-class sleep stage classification was proposed in \cite{zhang2017sleep}, using heart rate and wrist actigraphy acquired by a wristband. They achieved the higher accuracies when the sizes of datasets were much smaller. Compared to the works mentioned above, our results accomplish the significant improvement on the prediction accuracy, and the robust performance on a large dataset. We will have a detailed comparison in Table~\ref{tab:result} in Section \ref{sec:eval}.

\section{Hardware and Methods}
\label{sec:methods}

\begin{figure*}[t]
	\centering
	\hspace{.1pt}
	\subfloat[Wake]{%
		\includegraphics[width=0.24\linewidth]{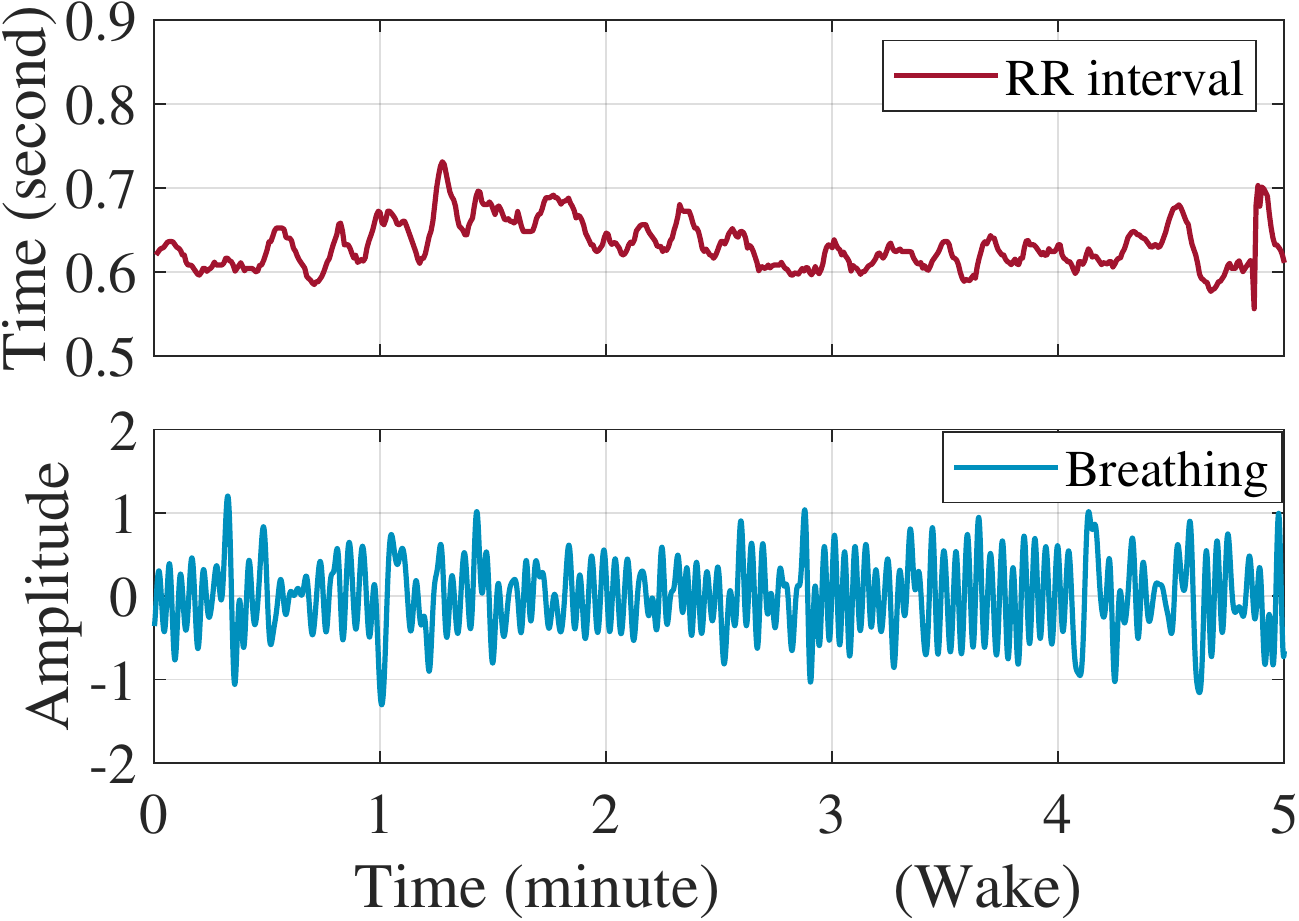}%
		\label{fig:wake}
	}
	\hfill
	\subfloat[Light sleep]{%
		\includegraphics[width=0.24\linewidth]{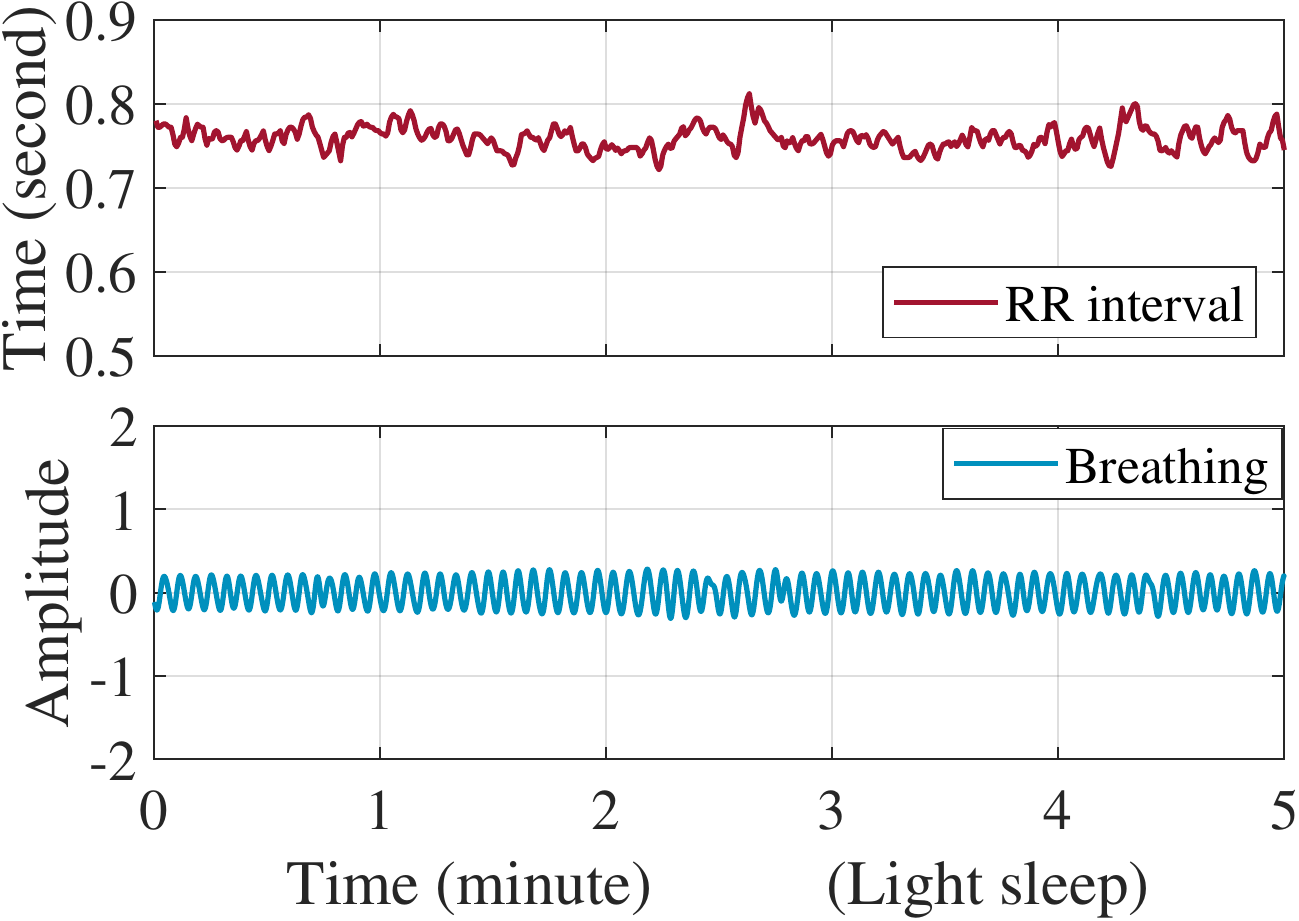}%
		\label{fig:light}
	}
	\hfill
	\subfloat[Deep sleep]{%
		\includegraphics[width=0.24\linewidth]{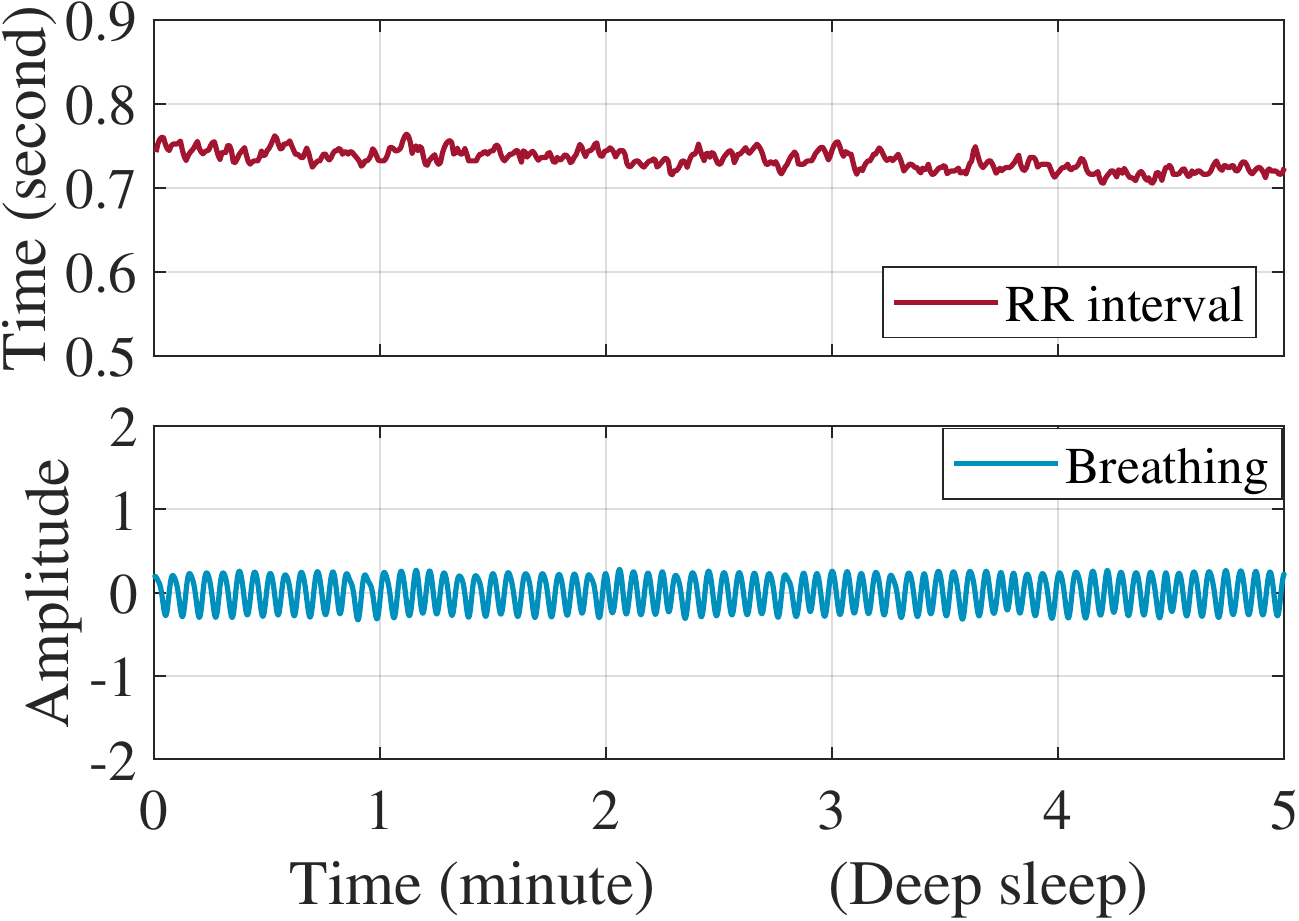}%
		\label{fig:deep}
	}
	\hfill	
	\subfloat[REM]{%
		\includegraphics[width=0.24\linewidth]{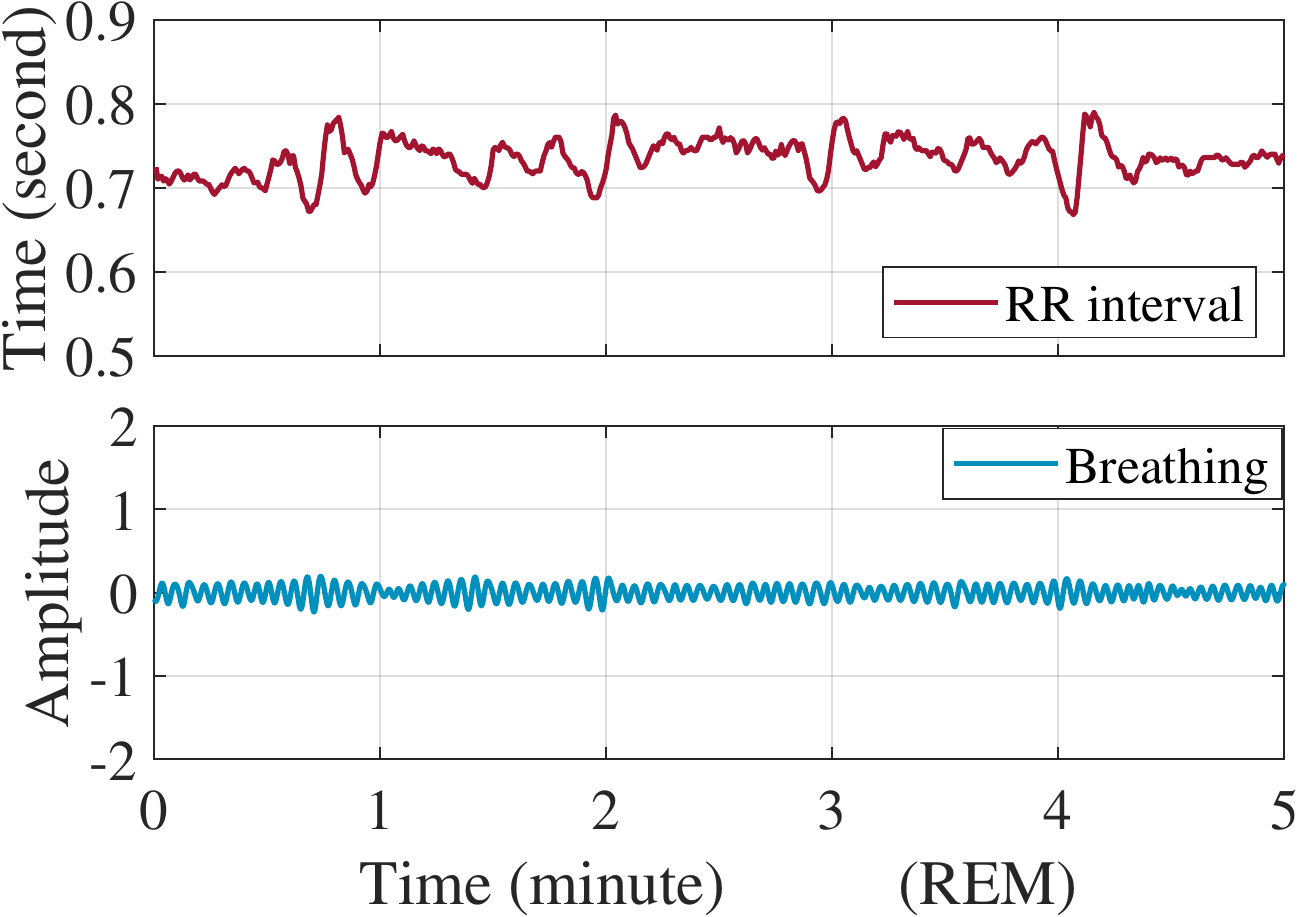}%
		\label{fig:rem}
	}
		\vspace{-5pt}
	\caption{\small{Examples of RR intervals and breathing signals in different sleep stages}}
	\label{fig:types}
	\vspace{-18pt}
\end{figure*}

\subsection{Hardware}
Our wearable multi-sensor system, SensEcho \cite{zhang2018automated,zhang2018breathing,lan2019tachy}, exploits three comfortable electrode patches to achieve the single-lead ECG signal monitoring at a 200 Hz sampling rate. Two sensing wires are embedded at the locations of chest and abdomen for the two corresponding breathing behaviors monitoring at a sampling rate of 25 Hz. In addition, an ultra-low power, 3-axis digital accelerometer component ADXL345 at a sampling rate of 25 Hz is also embedded in the vest. The main board of SensEcho is shown in Fig.~\ref{fig:main-board-smaller}. Fig.~\ref{fig:signal-overview} presents an example fragment to show the signal quality of our system compared to the PSG records. A wrist oximeter wirelessly communicates with SensEcho via Bluetooth, whose sampling rate is 1 Hz. SensEcho provides the local and secure cloud data storage options 
The 2GB-capacity local storage option enables us to backtrack the raw data when cloud storage is neither stable nor available. The embedded battery power can support continuous monitoring for five consecutive days.

\subsection{Data Pre-processing}
\label{sec:rr-intervals}
\textbf{ECG signals:} Rather than tackling the raw ECG signals, we extract \emph{RR intervals} from ECG signals.
RR interval is the time duration between two adjacent R-peaks of ECG signals
It can be used to calculate heart rate as well as Heart Rate Variability (HRV).

\textbf{Breathing signals:} As breathing signals always have the offset from the zero line of amplitude, we apply wavelet decomposition technologies to remove the offset effects. A Butterworth low-pass filter with the frequency of 1 Hz is next applied to remove signal noises at high frequency.

Fig.~\ref{fig:types} presents that the RR intervals and breathing signals on a subject along with the four-class sleep stage determination, respectively. It is obvious that the wake and the REM stages have distinguishing patterns on both RR intervals and breathing signals, while the light and deep sleep stages look similar to each other but minute differences on the RR intervals can be still found. Note that a specialist can easily distinguish these two sleep stages relying on the information from EEG, EOG, and EMG.

\subsection{Feature Extraction}
\label{sec:feature-extraction}

In this section, we elaborate our process of feature extraction. As a sleep stage is determined based on every 30-second \emph{epoch} in PSG tests \cite{quan1997sleep}, we define our sliding processing window unit as the same size. This way enables our time-series data to align the sleep stage labels in PSG records. Our feature extraction is processed on either one sliding window unit (one epoch), or a larger window consisting of several consecutive epochs. The moving step size is defined as one epoch. Note that when multiple consecutive epochs (a larger sliding window) are needed, the amount of epochs is an odd number such that the calculated feature would be exactly associated with the middle epoch. 
For example, extracting a feature from 3 consecutive epochs means that a feature is calculated inside a 90-second window and assigned to the second epoch.

\begin{table*}[t]
	\caption{\small{Comparison of four-class sleep stage classification (\textit{BE}: breathing effort; \textit{BLD}: Bayesian linear discriminant; \textit{BM}: body movement; \textit{BT}: boosted trees; \textit{CNN}: convolutional Neural Network; \textit{CRF}: conditional random field; \textit{GAN}: generative adversarial network; \textit{GBM}: gradient boosting machine; \textit{HB}: heart beat; \textit{RandF}: random forest; \textit{RR}: RR interval; \textit{SKNN}: subspace KNN; \textit{SVM}: support vector machine)}}
	\vspace{-5pt}
	\begin{tabular}{cccccccc}
		\hline
		\textbf{Authors} & \textbf{Data type} & \textbf{Dataset source} & \textbf{Data acquisition} & \textbf{Number of subjects} & \textbf{Method} & \textbf{Accuracy} & \Large{$\kappa$} \\
		\hline
		\cite{fonseca2015sleep} & RR+BE & SIESTA\cite{klosh2001siesta}+Own & PSG & 15+27 & BLD & 69.0\% & 0.49 \\
		\cite{tataraidze2016sleep} & RR+BE & SHHS & PSG & 625 & GBM+RandF & 71.4\% & 0.57 \\
		\cite{aggarwal2018sleep} & BE & MESA\cite{bild2002multi} & Ventilator & 400 & Deep CRF & 74.1\% & 0.57 \\
		\cite{muzet2016assessing} & RR+BM & Own & PSG & 48 & SVM & 74.9\% & n/a \\
		\cite{zhao2017learning} & HB+BE & Own & RF signals & 25 & CNN+RNN+GAN & 79.8\% & 0.70 \\
		\cite{hong2018noncontact} & HB+BE+BM & Own & RF signals & 13 & BT, SKNN, etc. & 81.0\% & n/a \\
		\multirow{4}{*}{\textbf{Ours}} & \multirow{4}{*}{\textbf{RR+BE}} & \multirow{4}{*}{\textbf{SHHS+Own}} & & \multirow{4}{*}{\textbf{417+32}} & \multirow{4}{*}{\textbf{BLSTM}} & \multicolumn{2}{c}{\textbf{Validation set from SHHS}}\\
		& & & \textbf{PSG (SHHS)} & & & \textbf{80.25\%} & \textbf{0.71} \\
		\cline{7-8}
		& & & \textbf{Wearable (own)} & & & \multicolumn{2}{c}{\textbf{Test set from own}} \\
		& & & & & & \textbf{80.75\%} & \textbf{0.69} \\
		\hline
	\end{tabular}
	\label{tab:result}
	\vspace{-15pt}
\end{table*}

\subsubsection{Features from RR intervals}
We extract numerous features from the time domain and the frequency domain on RR intervals.

\textbf{Time domain features:} We extract 10 features commonly used for the HRV analysis \cite{redmond2007sleep}, and 34 conventional statistical features on RR intervals \cite{fonseca2015sleep}, such as mean, quantiles, range, etc. We also extract 5 non-linear features including sample entropy, zero crossing analysis \cite{xiao2013sleep}. Nevertheless, the sudden changes of RR intervals are not well captured by using these features only. To address this issue, we design three novel features as follows:
\begin{equation}
f_1 = \overline{R^{mid}_n} - \overline{R_n}
\label{eq:f1}
\end{equation}
\vspace{-10pt}
\begin{equation}
f_2 = \overline{R^{mid}_n} - \widetilde{R_n}
\label{eq:f2}
\end{equation}
	\vspace{-10pt}
\begin{equation}
f_3 = \sqrt{\frac{1}{n}\sum_{i=1}^{n}(\overline{R^i_n} - \overline{R_n})^2}
\label{eq:f3}
\end{equation}
where $R_n$ represents raw RR intervals in the consecutive $n$ epochs; $R^i_n$ denotes the $i$th epoch of $R_n$; $R^{mid}_n$ refers to the middle epoch of $R_n$; $\overline{R_n}$ represents the average value of $R_n$; $\widetilde{R_n}$ denotes the median value of $R_n$. Such three features investigate the impact of sudden variation of RR intervals in one epoch over the longtime series. In practice, we set $n=119$ in $f_1$ to approximate one-hour-long time series, and set $n=9$ in $f_2$ and $f_3$ to consider 9 consecutive epochs (4.5 minutes). 

\textbf{Frequency domain features:} In order to obtain the frequency domain information, we first perform the linear interpolation on RR intervals, and then use Fast Fourier Transformation (FFT) on the given epoch(s). According to \cite{xiao2013sleep}, 21 features of the frequency domain are extracted, such as mean, spectrum power, entropy, etc.

\subsubsection{Features from breathing signals}
Similar to the feature extraction in RR intervals, we extract 25 statistical features from breathing signals according to \cite{redmond2007sleep,fonseca2015sleep}. 
For example, in the time domain, mean and standard deviance of breathing peak sequence, kurtosis, skewness, etc. are extracted; features in the frequency domain include dominant peak, energy, etc.

\subsubsection{Features from cardiopulmonary coupling}
Cardiopulmonary coupling (CPC) is a technology to analyze the connection between HRV and breathing volume variability. CPC index \cite{thomas2005electrocardiogram} is a set of magnitudes in the frequency domain to present CPC degrees at a given time spot. We calculate CPC indexes in $R_n$ and obtain the summation of CPC indexes at the high-frequency (HF) band (0.1--0.4Hz), at the low-frequency (LF) band (0.01--0.1Hz), and at the very-low-frequency (VLF) band (0--0.01Hz), respectively. The ratios of these summations at the specific frequency bands over the CPC index summation across the entire frequency domain are then calculated.

We finally extract 152 features from RR intervals, breathing signals, and cardiopulmonary coupling effects. At last, the Z-score normalization is performed on each feature such that its mean value is 0 and its variance value is 1. The feature vectors are fed into our learning model architecture. We then leverage two BLSTM layers, each of which has 16 units, and one 4-unit output layer corresponding to the four sleep stage classes. The advantage of BLSTM over conventional LSTM is that the BLSTM architecture is able to learn the target's both previous and future information.

\section{Evaluations}
\label{sec:eval}

\begin{figure*}[t]
	\centering
	\subfloat[Experiment set-up]{%
		\includegraphics[width=0.18\linewidth]{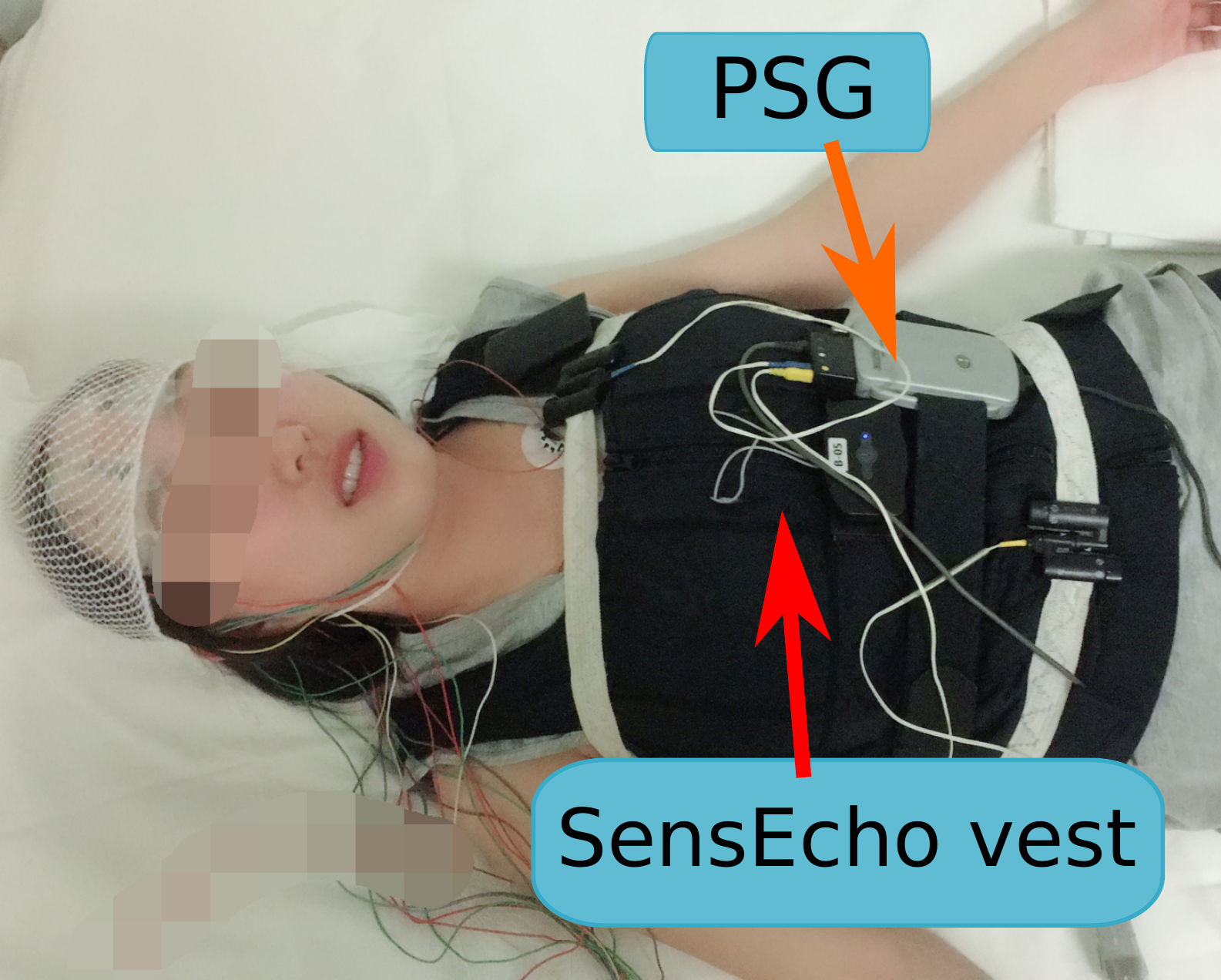}%
		\label{fig:zhangyuan}
	}
	\hfill
	\subfloat[Four-class confusion matrix on the validation set of the SHHS dataset]{%
		\includegraphics[width=0.185\linewidth]{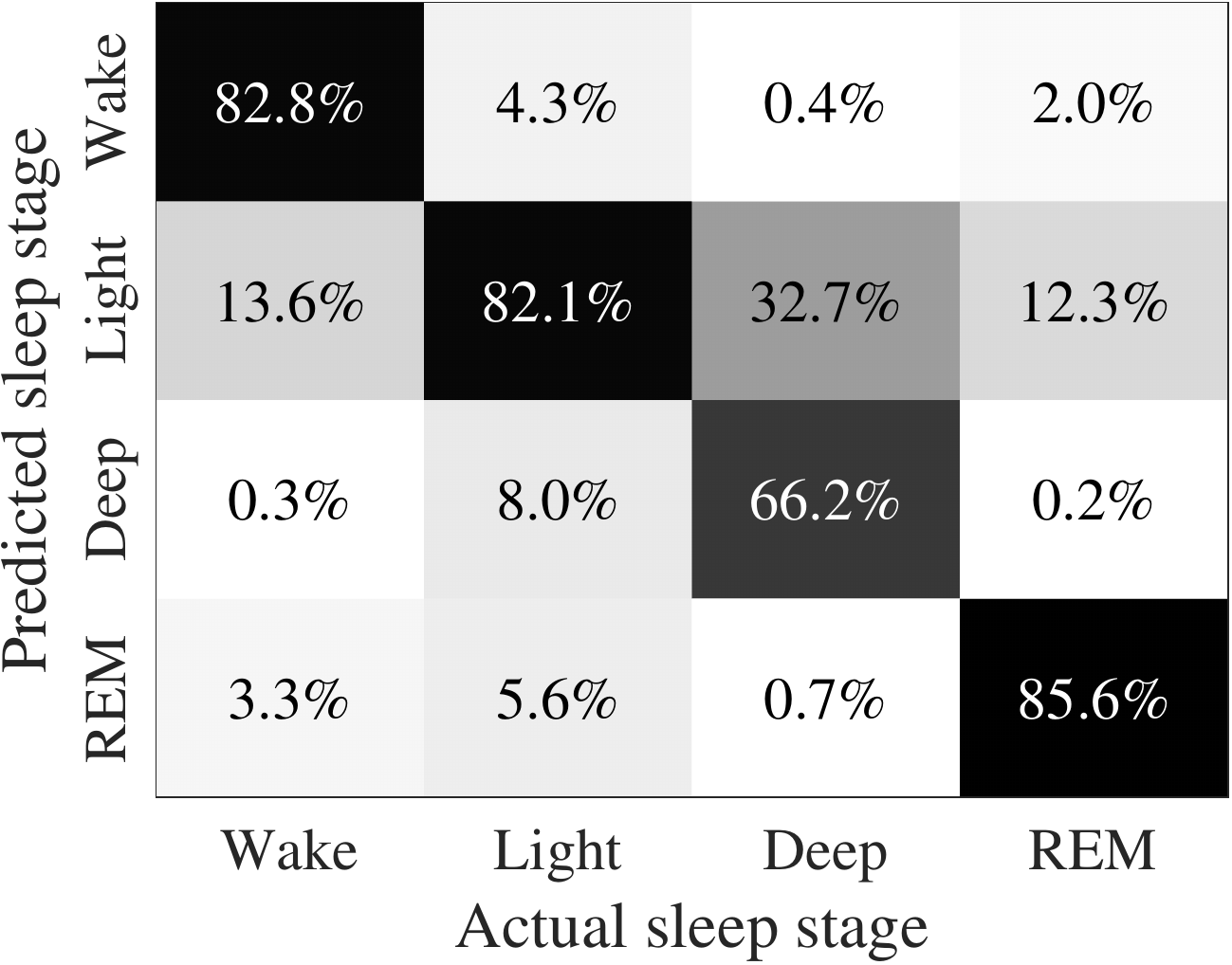}%
		\label{fig:confmat-SHHS}
	}
	\hfill
	\subfloat[Four-class confusion matrix on the test set (our own dataset)]{%
		\includegraphics[width=0.185\linewidth]{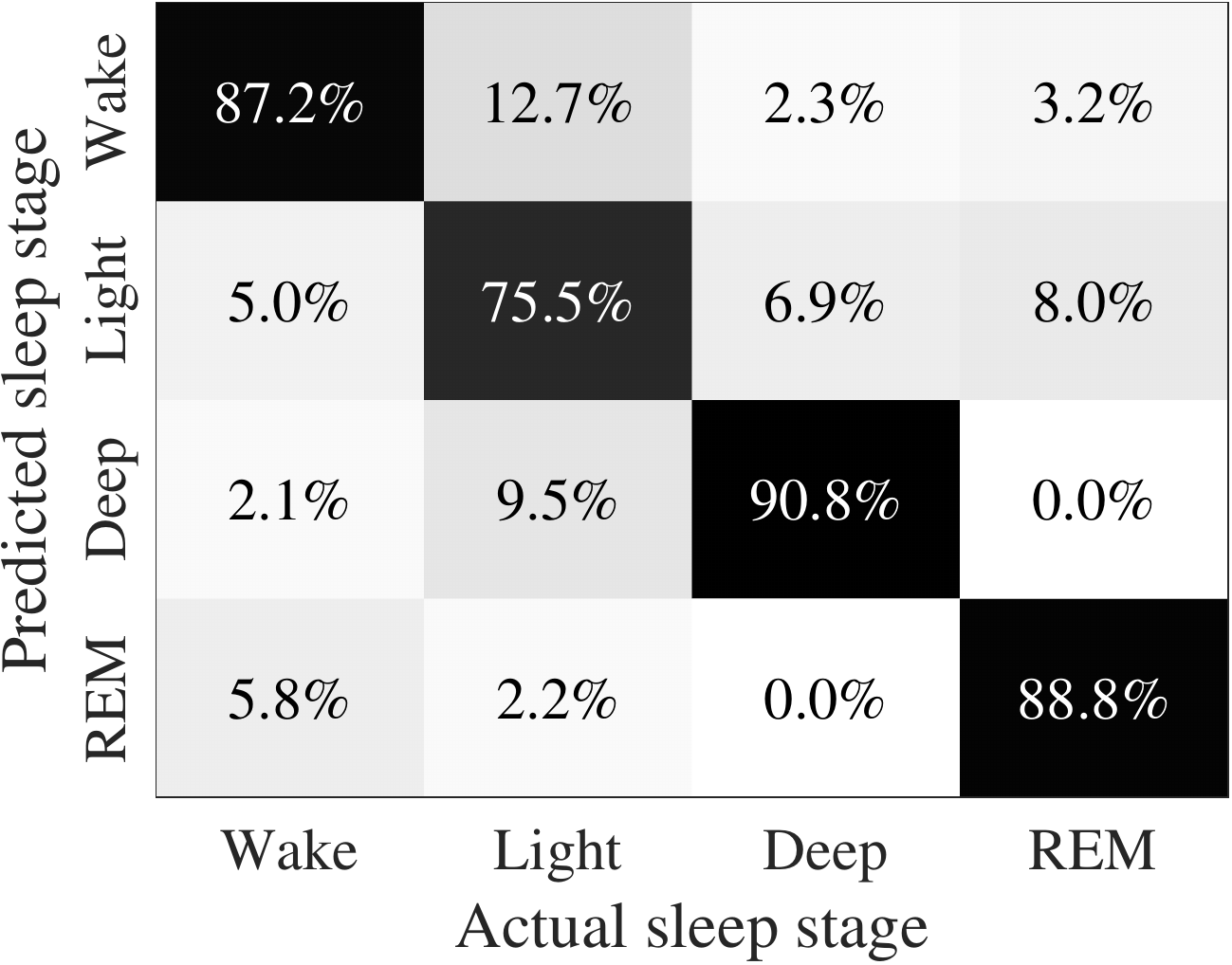}%
		\label{fig:confmat-ours}
	}
	\hfill	
	\subfloat[CDF of the prediction accuracies in different datasets]{%
		\includegraphics[width=0.18\linewidth]{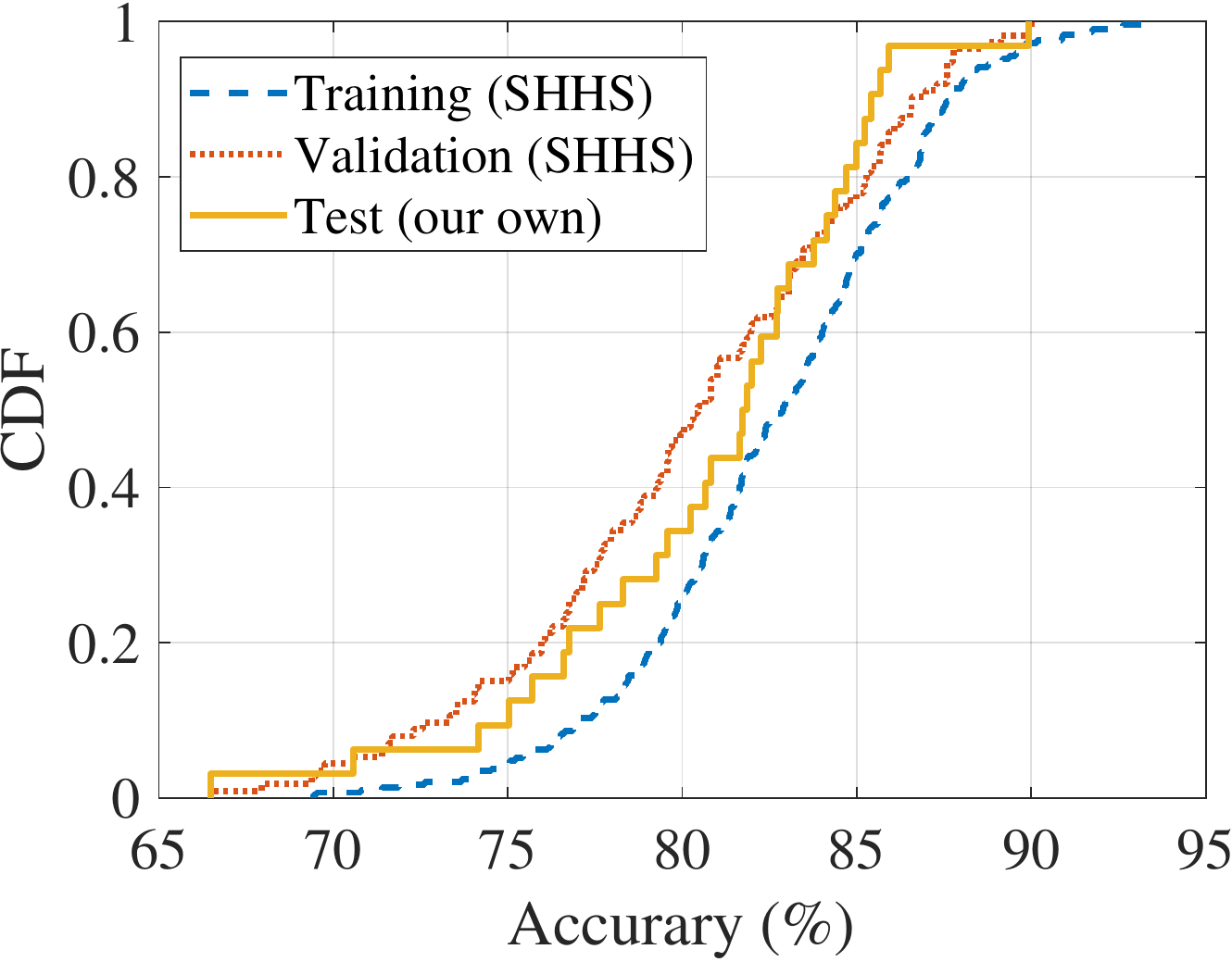}%
		\label{fig:cdf}
	}
	\hfill	
	\subfloat[Visualization of sleep epochs using t-SNE]{%
		\includegraphics[width=0.185\linewidth]{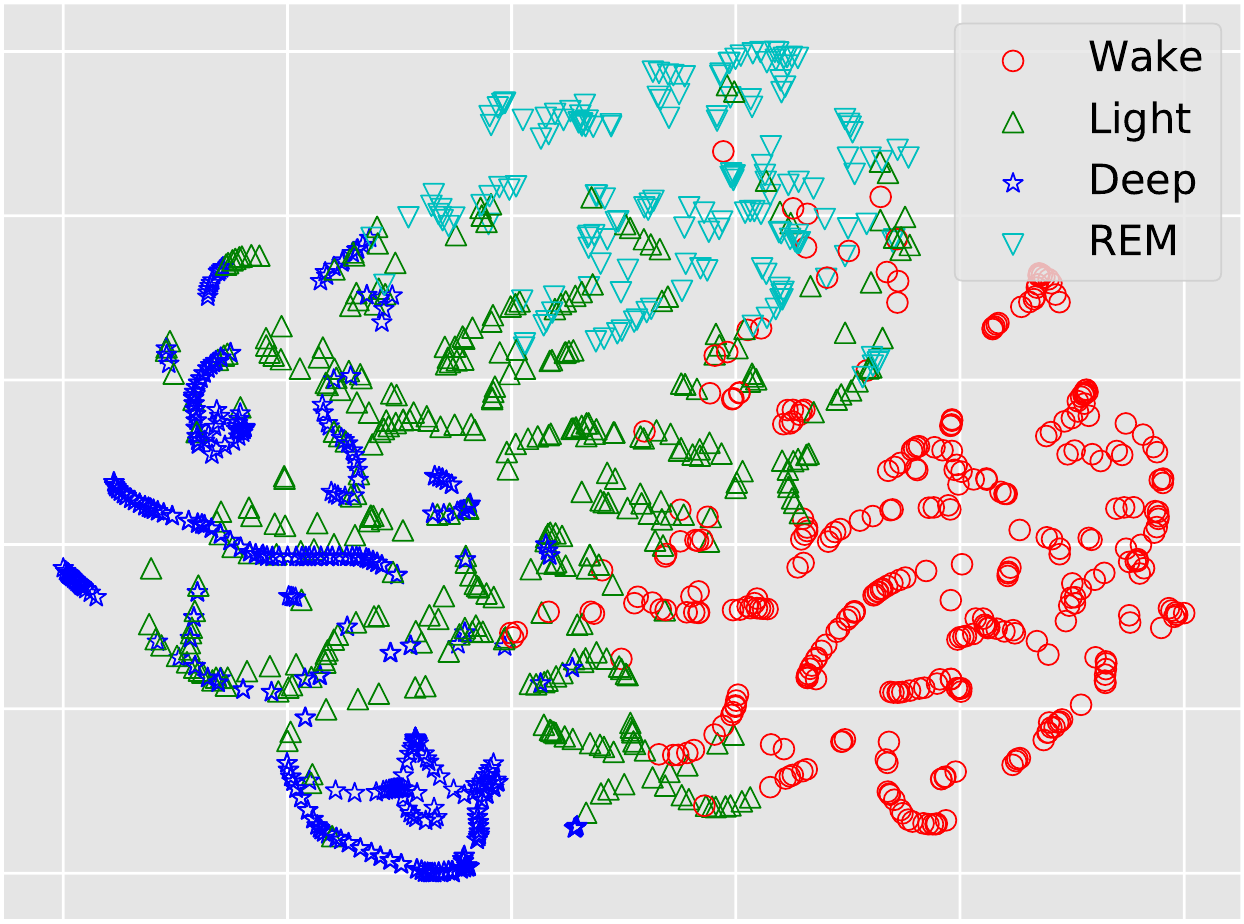}%
		\label{fig:tsne}
	}
		\vspace{-5pt}
	\caption{\small{Experiment set-up and visualization of results}}
	\label{fig:results}
		\vspace{-15pt}
\end{figure*}

\subsection{Datasets}
\subsubsection{Dataset from the public database}
To mitigate the training over-fitting issue using a small dataset, we leverage the large public sleep PSG database \emph{Sleep Heart Health Study} (SHHS) \cite{quan1997sleep} to build our training model. SHHS consists of the PSG monitoring during sleep of more than 6,000 subjects (52.4\% females) in the U.S. across the age range of $\sim$40+ years, including the records of six-class sleep stage classification for each subject manually determined by clinical specialists (wake, REM, S1, S2, S3, S4). It is worth noting that in the SHHS database numerous subjects have various sleep-related diseases, such as breathing disorders or irregularities, insomnia, etc. Those subjects would introduce serious bias to the training model. To minimize these impacts, we filter the data based on the following steps.
\begin{enumerate}
	\item We exploit the metric Apnea Hypopnea Index (AHI) score to indicate subjects' apnea levels, and further help to select the appropriate cohort for the training model. AHI score is calculated by the summation of manually identified breathing irregularity events. It could be commonly divided into four levels \cite{nandakumar2015contactless}: \emph{No apnea}: AHI $<$ 5; \emph{Mild}: 5 $\leq$ AHI $<$ 15; \emph{Medium}: 15 $<$ AHI $\leq$ 30; and \emph{Severe}: AHI $>$ 30. The subjects who have no apnea (AHI $<$ 5) are chosen as candidates to build our data set from SHHS.
	
	\item Among the subject candidates, we then select the subjects who have at least 5\% of S3 and S4 stages, and at least 15\% REM stage, during their sleep periods, since sleep with such quality is regarded as \emph{regular sleep} \cite{fonseca2017cardiorespiratory}
	
	\item We combine the S1 and S2 stages as the light sleep stage, and combine the S3 and S4 stages as the deep sleep stage \cite{berry2012aasm}. With the existing stages of wake and REM, the six-class sleep stages are converted into four-class ones.
	
	\item As a result, 417 subjects are selected to construct the dataset from the SHHS database. We randomly take 70\% of them for the training set, while the remaining 30\% contribute to the validation set. The subjects constituting the training set never appear in the validation set.
\end{enumerate}

\begin{figure}[t]	
	\begin{center}

		\includegraphics[width=\linewidth]{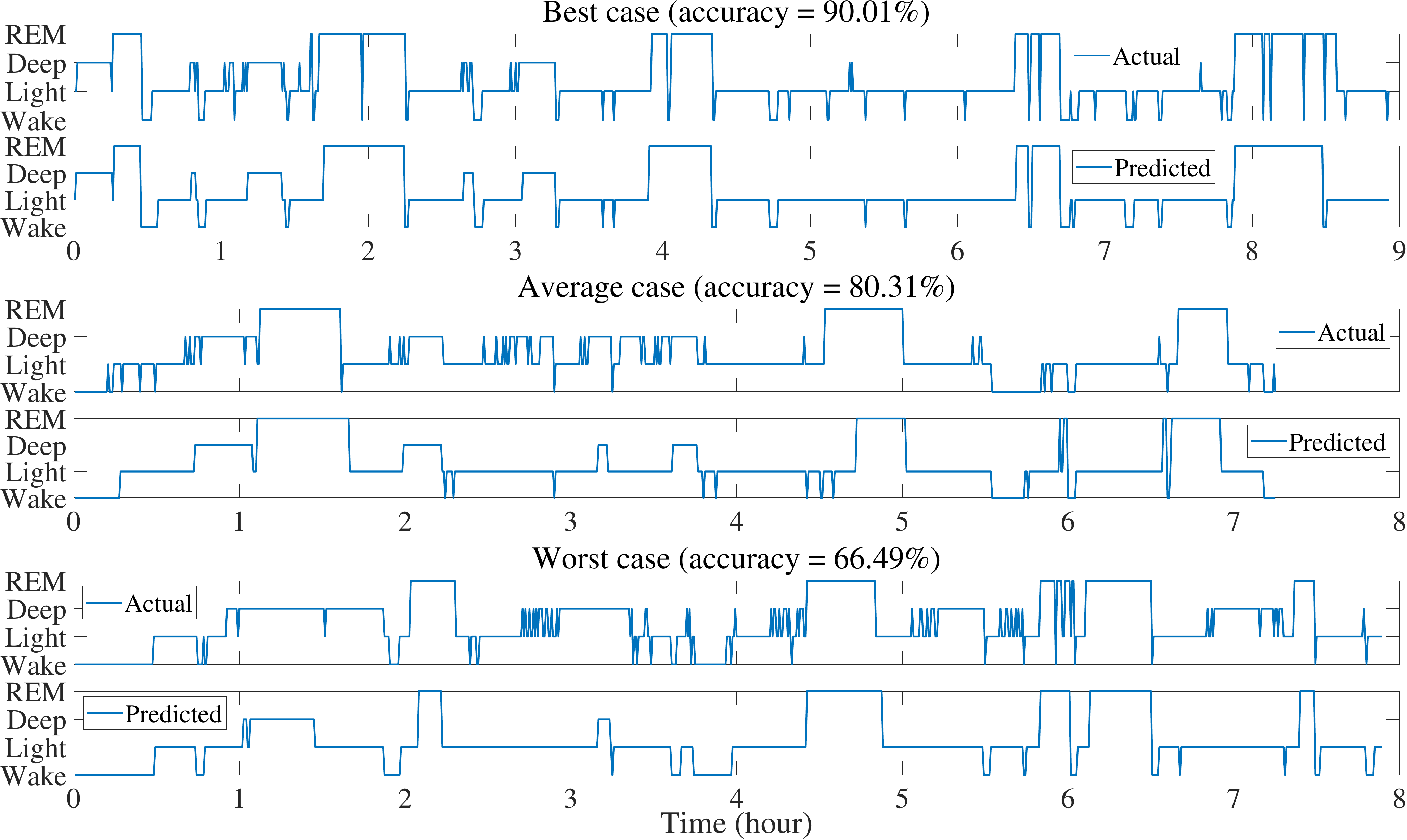}
				\vspace{-13pt}
		\caption{\small{Examples of best, average, worst classifications in the validation set of SHHS (referring to the validation set in Fig.~\ref{fig:cdf})}}
		\label{fig:three-cases}
				\vspace{-20pt}
	\end{center}
\end{figure}

\subsubsection{Our own dataset}
Aside from using the public database, we are also building our own dataset for sleep stage classification. Till now, 32 subjects (22 -- 45 ages and 15 females) with no apnea (AHI $<$ 5) are enrolled to wear SensEcho during the PSG tests at Chinese PLA General Hospital (shown in Fig.~\ref{fig:zhangyuan}). 
Their four-class sleep stages are identified by three clinical specialists' consent. Since the size of the dataset is now still limited, it is served as the test set only. In other words, we leverage the trained model from the SHHS training set described above to make the prediction on our own dataset. Note that this study is reviewed and approved by the Ethics Committee of Chinese PLA General Hospital (IRB number: S2018-095-01)

\subsubsection{Consistency of the datasets}
As aforementioned in Sec.~\ref{sec:feature-extraction}, we align our data with the sleep stage determination based on the standard PSG records using the 30-second epoch. Such processing approach guarantees that the model trained from SHHS dataset can be applied to predict the test set (i.e., our own dataset).

\subsection{Sleep Stage Classification}

Table \ref{tab:result} lists the recent literature which contributes to the four-class sleep stage classification using RR interval/heart rate and breathing effort, where some works also involve the information of body movements. As we see, the sizes of datasets used in their works are very diverse. Compared to the previous works using large dataset \cite{tataraidze2016sleep,aggarwal2018sleep}, the BLSTM method in our work significantly outperform them. The reason is that BLSTM is able to learn the hidden relationship in time-series correlated signals. Meanwhile, aside from overcoming the over-fitting problem for the model training, we keep the similar accuracy level compared to the works which use small data sets to build the learning model \cite{zhao2017learning,hong2018noncontact}. 
Note that we let \Large{$\kappa$} \normalsize denote \emph{Cohen's kappa coefficient}, which is a more robust metric than the accuracy percentage as it considers the occurrence by chance \cite{cohen1960coefficient}.

Fig.~\ref{fig:confmat-SHHS} presents that the confusion matrix of four sleep stage classes on the validation set of the SHHS dataset. We observe that except deep sleep, all other three stages achieve great classification accuracies. However, plenty of deep sleep epochs are classified into light sleep epochs. This observation aligns with the examples of these two stages in Figs.~\ref{fig:light} and \ref{fig:deep}, where the patterns of RR intervals and breathing signals are difficult to distinguish. We then apply the prediction model to our test set to obtain another confusion matrix result (shown in Fig.~\ref{fig:confmat-ours}). While the high prediction accuracies of wake and REM stages are maintained, it is interesting that the accuracy of deep sleep increases much but the accuracy of light sleep declines. It is worth recalling the different age ranges of the training and validation sets and the test set. As observed in \cite{peters2014age}, the light sleep of young subjects is less than it of senior subjects, while the deep sleep of young subjects is more. Furthermore, the ethnic differences (White in SHHS vs. Asian in our dataset) and gender distributions also lead to dissimilar sleep patterns \cite{rao2009ethnic}. Due to these reasons, we believe that there is some over-fitting issue during the light sleep stage training but that the younger people's deep sleep patterns are clearer to be identified and predicted by the prediction model.
Fig.~\ref{fig:cdf} shows that the Cumulative Distribution Function (CDF) of the accuracies of the three sets in terms of subjects. Even though there is some generational bias in the training model, the overall prediction accuracy on the test set still maintains the same level as it on the validation set. The smaller size of test set might result in the slightly better prediction result.

\begin{figure}[t]
	\centering
	\includegraphics[width=\linewidth]{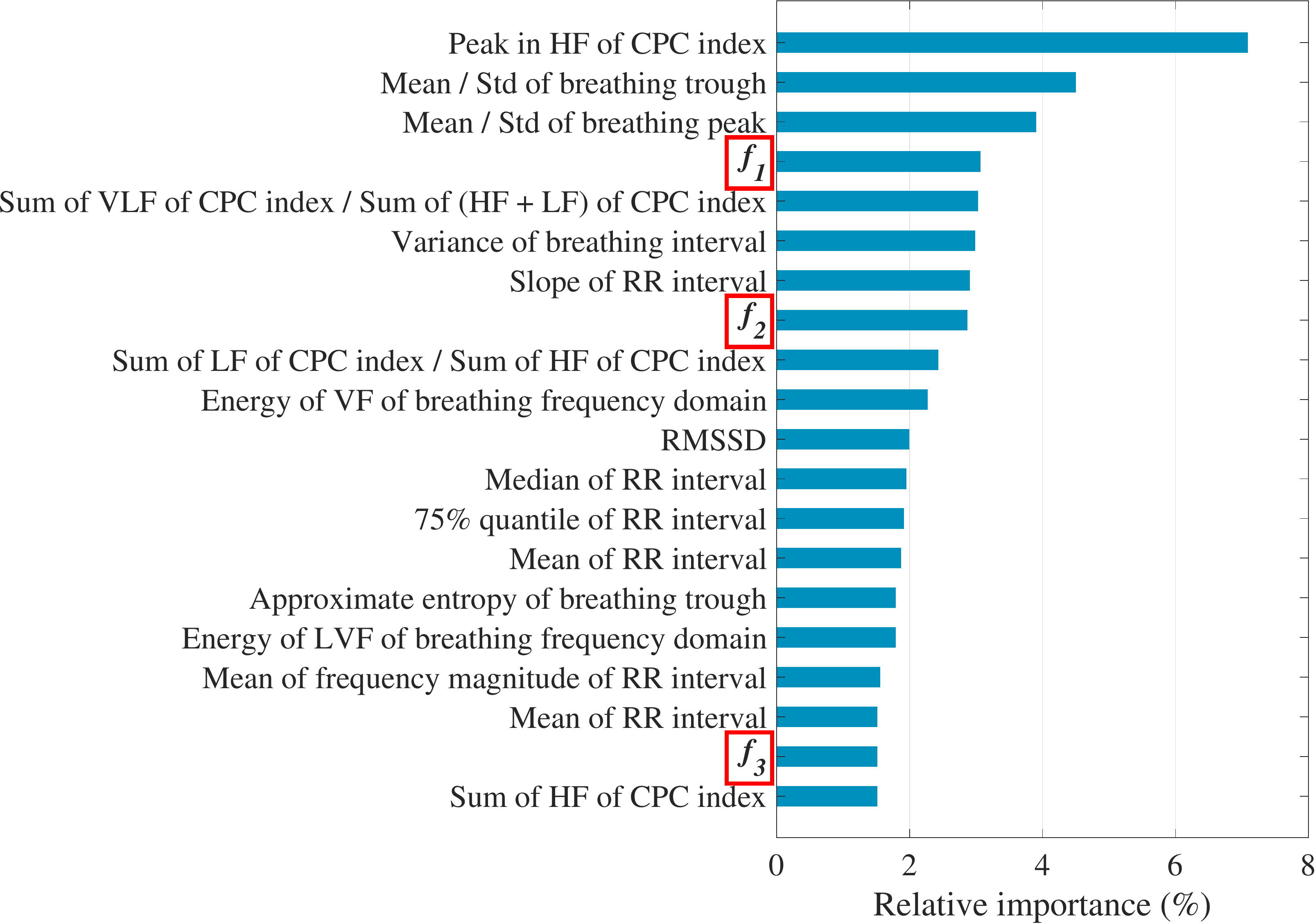}%
	\caption{\small{Top 20 significant features ranked in XGBoost. (\textit{RMSSD}: root mean square of the successive differences}}
	\label{fig:feature-importance}
	\vspace{-10pt}
\end{figure}

We then select the best case, the worst case, and one average case from the validation set to view the prediction details. Fig.~\ref{fig:three-cases} shows that the major impact on the prediction accuracy is the transition between the light sleep stage and the deep sleep stage, which is not well recognized, especially in the case that such transitions occur many times during the sleep period (i.e. the worst case). However, the transition between the light sleep stage and the wake stage is well predicted across these three cases. Considering the distinct difference in RR intervals and breathing signals of these two stages in Figs.~\ref{fig:wake} and \ref{fig:light}, our three novel features are designed for capturing such a sudden transition even though it lasts for only one epoch. We will evaluate them in the next section.

\subsection{Knowledge of Features}

In this section, we investigate how the extracted features contribute to the learning process. Fig.~\ref{fig:tsne} visualizes the sleep epochs of one subject using the t-SNE \cite{maaten2008visualizing}. We notice that the system is able to learning much coherent information across the four classes. The interweaving between the deep sleep epochs and the light sleep epochs are more than all other two stage pairs. Since it is still difficult to interpret the significances of specific features, we leverage the recent common used tree model, XGBoost \cite{chen2016xgboost}, to sort all the features' relative importance according to their \emph{weights}. The weight refers to the times a feature is used to split the data across all the trees. Fig.~\ref{fig:feature-importance} lists the top 20 important features which contribute to the classification most significantly. It is worth noting that our three designed features $f_1$, $f_2$, and $f_3$ are ranked at No. 4, 8, and 19, respectively. We also observe that several features extracted from CPC are top significant, indicating the indispensable synergistic effects of cardiorespiratory signals on different sleep stages. 
Note that the prediction accuracy using XGBoost is 75.6\% on the validation set.

\section{Discussions, Potentials And Future Work}
\label{sec:discussion}

As we have shown that our approach compared to the latest contributions has an outstanding sleep stage classification accuracy on a large dataset, there are many potential directions in our future works to further pursue a better performance. Authors in \cite{zhao2017learning} leverage CNN to create effective features which are difficult interpreted and extracted manually. We are currently improving the BLSTM model and designing an appropriate CNN architecture to integrate with our current feature sets. Meanwhile, alongside recruiting healthy subjects, we are enrolling actual patients using our wearable SensEcho who have heart, breathing, or sleep issues/diseases, in order to build our comprehensive sleep datasets for the intra-cohort and inter-cohort sleep researches. As aforementioned, a 3-axis accelerometer is embedded in SensEcho which can infer the body movement behaviors in previous works \cite{muzet2016assessing,hong2018noncontact}. Our future works will also involve the use of accelerometer data.

\section{Conclusions}
\label{sec:conclusion}
In this paper, we present our four-class sleep stage classification trained by two BLSTM layers. We use our low-cost wearable multi-sensor system, SensEcho, to acquire subjects' RR intervals and breathing signals. Three novel features are designed during the feature extraction to detect the sudden variation on RR intervals. Our robust prediction accuracy is 80.25\% on a large public dataset (417 subjects), and 80.75\% on our 32 enrolled subjects, respectively. These results are better than the previous works which either used small data sets with the potential over-fitting issues, or used the conventional machine learning methods on large data sets.

\section*{Acknowledgments}
This work was supported by NSF of China (61471398), Beijing Municipal Science and Technology (Z181100001918023), CERNET Innovation Project (NGII20160701), Translational Medicine Project of Chinese PLA General Hospital (2016TM-041) and Big Data Research \& Development Project of Chinese PLA General Hospital (2018MBD-09).

\bibliographystyle{IEEEtran}
\bibliography{IEEEabrv,bib} 

\begin{thebibliography}{10}
\providecommand{\url}[1]{#1}
\csname url@samestyle\endcsname
\providecommand{\newblock}{\relax}
\providecommand{\bibinfo}[2]{#2}
\providecommand{\BIBentrySTDinterwordspacing}{\spaceskip=0pt\relax}
\providecommand{\BIBentryALTinterwordstretchfactor}{4}
\providecommand{\BIBentryALTinterwordspacing}{\spaceskip=\fontdimen2\font plus
\BIBentryALTinterwordstretchfactor\fontdimen3\font minus
  \fontdimen4\font\relax}
\providecommand{\BIBforeignlanguage}[2]{{%
\expandafter\ifx\csname l@#1\endcsname\relax
\typeout{** WARNING: IEEEtran.bst: No hyphenation pattern has been}%
\typeout{** loaded for the language `#1'. Using the pattern for}%
\typeout{** the default language instead.}%
\else
\language=\csname l@#1\endcsname
\fi
#2}}
\providecommand{\BIBdecl}{\relax}
\BIBdecl

\bibitem{berry2012aasm}
R.~B. Berry, R.~Brooks, C.~E. Gamaldo \emph{et~al.}, ``The aasm manual for the
  scoring of sleep and associated events,'' \emph{Rules, Terminology and
  Technical Specifications, Darien, Illinois, American Academy of Sleep Med},
  2012.

\bibitem{yang2017vital}
Z.~Yang, P.~H. Pathak, Y.~Zeng, X.~Liran, and P.~Mohapatra, ``Vital sign and
  sleep monitoring using millimeter wave,'' \emph{ACM TOSN}, 2017.

\bibitem{dor2012experiences}
R.~Dor, G.~Hackmann, Z.~Yang, C.~Lu, Y.~Chen, M.~Kollef, and T.~Bailey,
  ``Experiences with an end-to-end wireless clinical monitoring system,'' in
  \emph{ACM Wireless Health}.\hskip 1em plus 0.5em minus 0.4em\relax ACM, 2012.

\bibitem{zhang2017sleep}
X.~Zhang, W.~Kou, E.~I. Chang, H.~Gao \emph{et~al.}, ``Sleep stage
  classification based on multi-level feature learning and recurrent neural
  networks via wearable device,'' \emph{arXiv:1711.00629}, 2017.

\bibitem{zhao2017learning}
M.~Zhao, S.~Yue, D.~Katabi, T.~S. Jaakkola, and M.~T. Bianchi, ``Learning sleep
  stages from radio signals: a conditional adversarial architecture,'' in
  \emph{ACM ICML}, 2017.

\bibitem{graves2005framewise}
A.~Graves and J.~Schmidhuber, ``Framewise phoneme classification with
  bidirectional lstm and other neural network architectures,'' \emph{Neural
  Networks}, vol.~18, no. 5-6, pp. 602--610, 2005.

\bibitem{fonseca2015sleep}
P.~Fonseca, X.~Long, M.~Radha, R.~Haakma, R.~M. Aarts, and J.~Rolink, ``Sleep
  stage classification with ecg and respiratory effort,'' \emph{Physiological
  measurement}, 2015.

\bibitem{muzet2016assessing}
A.~Muzet, S.~Werner, G.~Fuchs \emph{et~al.}, ``Assessing sleep architecture and
  continuity measures through the analysis of heart rate and wrist movement
  recordings in healthy subjects: comparison with results based on
  polysomnography,'' \emph{Elsevier Sleep medicine}, 2016.

\bibitem{hong2018noncontact}
H.~Hong, L.~Zhang, C.~Gu, Y.~Li, G.~Zhou, and X.~Zhu, ``Noncontact sleep stage
  estimation using a cw doppler radar,'' \emph{IEEE JETCAS}, 2018.

\bibitem{tataraidze2016sleep}
A.~Tataraidze, L.~Korostovtseva, L.~Anishchenko, M.~Bochkarev, and Y.~Sviryaev,
  ``Sleep architecture measurement based on cardiorespiratory parameters,'' in
  \emph{IEEE EMBC}, 2016.

\bibitem{aggarwal2018sleep}
K.~Aggarwal, S.~Khadanga, S.~R. Joty, L.~Kazaglis, and J.~Srivastava, ``Sleep
  staging by modeling sleep stage transitions using deep crf,'' \emph{arXiv
  preprint arXiv:1807.09119}, 2018.

\bibitem{pan2012transition}
S.-T. Pan, C.-E. Kuo, J.-H. Zeng, and S.-F. Liang, ``A transition-constrained
  discrete hidden markov model for automatic sleep staging,'' \emph{Biomedical
  engineering online}, 2012.

\bibitem{supratak2017deepsleepnet}
A.~Supratak, H.~Dong, C.~Wu, and Y.~Guo, ``Deepsleepnet: a model for automatic
  sleep stage scoring based on raw single-channel eeg,'' \emph{IEEE TNSRE},
  2017.

\bibitem{zhang2018automated}
Y.~Zhang, Z.~Yang, Z.~Zhang, X.~Liu, D.~Cao, P.~Li, J.~Zheng, and K.~Lan,
  ``Automated sleep period estimation in wearable multi-sensor systems,'' in
  \emph{Proceedings of the 16th ACM Conference on Embedded Networked Sensor
  Systems}.\hskip 1em plus 0.5em minus 0.4em\relax ACM, 2018.

\bibitem{zhang2018breathing}
Y.~Zhang, Z.~Yang, Z.~Zhang, P.~Li, D.~Cao, X.~Liu, J.~Zheng, Q.~Yuan, and
  J.~Pan, ``Breathing disorder detection using wearable electrocardiogram and
  oxygen saturation,'' in \emph{Proceedings of the 16th ACM Conference on
  Embedded Networked Sensor Systems}.\hskip 1em plus 0.5em minus 0.4em\relax
  ACM, 2018.

\bibitem{lan2019tachy}
K.~Lan, X.~Liu, H.~Xu, P.~Li, Z.~Yang, Q.~Yuan, J.~Zheng, W.~Yan, D.~Cao, and
  Z.~Zhang, ``Deeptop: Personalized tachycardia onset prediction using
  bi-directional lstm in wearable embedded systems,'' in \emph{2019
  International Conference on Embedded Wireless Systems and Networks
  (EWSN)}.\hskip 1em plus 0.5em minus 0.4em\relax ACM, 2019.

\bibitem{quan1997sleep}
S.~F. Quan, B.~V. Howard, C.~Iber \emph{et~al.}, ``The sleep heart health
  study: design, rationale, and methods,'' \emph{Sleep}, 1997.

\bibitem{klosh2001siesta}
G.~Klosh, B.~Kemp, T.~Penzel \emph{et~al.}, ``The siesta project polygraphic
  and clinical database,'' \emph{IEEE Engineering in Medicine and Biology
  Magazine}, 2001.

\bibitem{bild2002multi}
D.~E. Bild, D.~A. Bluemke, G.~L. Burke, R.~Detrano, A.~V. Diez~Roux
  \emph{et~al.}, ``Multi-ethnic study of atherosclerosis: objectives and
  design,'' \emph{American Journal of Epidemiology}, 2002.

\bibitem{redmond2007sleep}
S.~J. Redmond, P.~de~Chazal, C.~O'Brien, S.~Ryan, W.~T. McNicholas, and
  C.~Heneghan, ``Sleep staging using cardiorespiratory signals,''
  \emph{Somnologie-Schlafforschung und Schlafmedizin}, 2007.

\bibitem{xiao2013sleep}
M.~Xiao, H.~Yan, J.~Song, Y.~Yang, and X.~Yang, ``Sleep stages classification
  based on heart rate variability and random forest,'' \emph{Biomedical Signal
  Processing and Control}, 2013.

\bibitem{thomas2005electrocardiogram}
R.~J. Thomas, J.~E. Mietus, C.-K. Peng, and A.~L. Goldberger, ``An
  electrocardiogram-based technique to assess cardiopulmonary coupling during
  sleep,'' \emph{Sleep}, 2005.

\bibitem{nandakumar2015contactless}
R.~Nandakumar, S.~Gollakota, and N.~Watson, ``Contactless sleep apnea detection
  on smartphones,'' in \emph{ACM Mobisys}, 2015.

\bibitem{fonseca2017cardiorespiratory}
P.~Fonseca, N.~den Teuling, X.~Long, and R.~M. Aarts, ``Cardiorespiratory sleep
  stage detection using conditional random fields,'' \emph{IEEE J. Biomed.
  Health Inf}, 2017.

\bibitem{cohen1960coefficient}
J.~Cohen, ``A coefficient of agreement for nominal scales,'' \emph{Educational
  and psychological measurement}, vol.~20, no.~1, pp. 37--46, 1960.

\bibitem{peters2014age}
K.~R. Peters, L.~B. Ray, S.~Fogel, V.~Smith, and C.~T. Smith, ``Age differences
  in the variability and distribution of sleep spindle and rapid eye movement
  densities,'' \emph{PLoS one}, 2014.

\bibitem{rao2009ethnic}
U.~Rao, C.~L. Hammen, and R.~E. Poland, ``Ethnic differences in
  electroencephalographic sleep patterns in adolescents,'' \emph{Asian J. of
  psychiatry}, 2009.

\bibitem{maaten2008visualizing}
L.~v.~d. Maaten and G.~Hinton, ``Visualizing data using t-sne,'' \emph{Journal
  of machine learning research}, 2008.

\bibitem{chen2016xgboost}
T.~Chen and C.~Guestrin, ``Xgboost: A scalable tree boosting system,'' in
  \emph{Proceedings of the 22nd acm sigkdd international conference on
  knowledge discovery and data mining}.\hskip 1em plus 0.5em minus 0.4em\relax
  ACM, 2016, pp. 785--794.

\end{thebibliography}

\end{document}